\documentclass[preprint,12pt]{elsarticle}

\usepackage{amsmath, amssymb, amsfonts, latexsym, bm, graphicx}
\newcommand{\bra}[1]{\langle #1 |}
\newcommand{\ket}[1]{|#1\rangle}
\newcommand{\braket}[2]{\langle #1 | #2 \rangle}
\newcommand{\ketbra}[2]{|#1 \rangle \langle #2|}

\newcommand{\expect}[1]{\langle #1 \rangle}
\newcommand{\abs}[1]{|#1|}

\begin{document}

\begin{frontmatter}

%% Title, authors and addresses

%% use the tnoteref command within \title for footnotes;
%% use the tnotetext command for the associated footnote;
%% use the fnref command within \author or \address for footnotes;
%% use the fntext command for the associated footnote;
%% use the corref command within \author for corresponding author footnotes;
%% use the cortext command for the associated footnote;
%% use the ead command for the email address,
%% and the form \ead[url] for the home page:
%%
%% \title{Title\tnoteref{label1}}
%% \tnotetext[label1]{}
%% \author{Name\corref{cor1}\fnref{label2}}
%% \ead{email address}
%% \ead[url]{home page}
%% \fntext[label2]{}
%% \cortext[cor1]{}
%% \address{Address\fnref{label3}}
%% \fntext[label3]{}

\title{Causality in quantum physics, the ensemble of beginnings of time, and the dispersion relations of wave function}

%% use optional labels to link authors explicitly to addresses:
%% \author[label1,label2]{<author name>}
%% \address[label1]{<address>}
%% \address[label2]{<address>}

\author{Y. Sato}
\ead{ysato@colby.edu}
\address{Department of Physics and Astronomy, Colby College, 5800 Mayflower Hill
\\Waterville, Maine 04901-8858}

\author{A. R. Bohm}
\ead{bohm@physics.utexas.edu}
\address{Department of Physics, University of Texas at Austin, 1 University Station C1600
Austin, Texas 78712-0264}

\begin{abstract}
%% Text of abstract
In quantum physics, disturbance due to a measurement is not negligible.
	This requires the time parameter $t$ in the Schr\"odinger or Heisenberg equation to be considered differently from a time continuum of experimenter's clock $T$ on which physical events are recorded.
	It will be shown that $t$ represents an ensemble of time intervals on $T$ during which a microsystem travels undisturbed.
	In particular $t=0$ represents the ensemble of preparation events that we refer to as the ensemble of beginnings of time.
	This restricts $t$ to be $0\leq t<\infty$. 
	But such a time evolution of quantum states cannot be achieved in the Hilbert space $L^2$ functions because due to the Stone-von Neumann theorem this time evolution is given by the unitary group with $t$ extending $-\infty<t<\infty$.
	Hence one needs solutions of the Schr\"odinger (and Heisenberg) equation under time asymmetric boundary condition in which only the semigroup time evolution is allowed.
	This boundary condition is fulfilled by the energy wave functions for quantum states (and as well as for observables) which are smooth Hardy function satisfying the Hilbert transform, called the dispersion relation in physics.

\end{abstract}

\begin{keyword}
%% keywords here, in the form: keyword \sep keyword

%% MSC codes here, in the form: \MSC code \sep code
%% or \MSC[2008] code \sep code (2000 is the default)
quantum mechanics, principle, causality
\end{keyword}

\end{frontmatter}

%%
%% Start line numbering here if you want
%%
% \linenumbers

%% main text
%\section{}
%\label{}

%%%%%%%%%%%%%%%%%%%%%%%%%%%%%%%%%%%%%%%%%%%%%%%%%%%%%%%%%%%%%
\section{Introduction}
%%%%%%%%%%%%%%%%%%%%%%%%%%%%%%%%%%%%%%%%%%%%%%%%%%%%%%%%%%%%%
	Nearly eighty years ago, Kronig~\cite{ref:kronig26} and Kramers~\cite{ref:kramers27} discovered a relation between causality and analyticity of a complex refractive index.
	They used the condition (causality) that a signal cannot travel faster than the speed of light, and the analyticity derived from it was a simple integral formula relating a dispersive process to an absorption process~\cite{ref:toll56}.
	This analyticity is generally referred to as dispersion relation.
	Kronig~\cite{ref:kronig42} then proved that the dispersion relation is the necessary and sufficient condition for the causality condition to be satisfied.
	Since then, the dispersion relation has been generalized and used in may branches of physics~\cite{ref:toll56, ref:nussenzveig72}.
	In non-relativistic quantum physics, Schutzer and Tiomno~\cite{ref:schutzer51}  derived a dispersion relation of a S-matrix element for the scattering of particles by a finite range scatterer; their causality condition was that the out-going scattered wave must be zero for all times before the in-going incident wave hits the scattering center.
%~\cite{ref:goldberger64}. 
	This work was criticized and generalized by van Kampen~\cite{ref:vankampen53II}, and its even more general derivation was given later by Wigner~\cite{ref:wigner55}.
	In relativistic quantum physics, Gell-Mann, Goldberger, and Thirring~\cite{ref:gell-mann54} derived a dispersion relation of S-matrix elements from their causality condition, called microscopic causality, that the commutator (or anticommutator for fermions) of two Heisenberg field operators taken at space-like points vanish.
	This has been provided a non-perturbative method for relativistic quantum field theory~\cite{ref:weinberg95}.
%	Thus all of the dispersion relations obtained so far in quantum physics are of S-matrix elements.

	Our article presents yet another dispersion relation in quantum physics, a dispersion relation that the wave functions satisfy in the energy representation.
	The causality condition we employ here is the quantum physics version of ``cause and effect'' discussed in detail by Ludwig~\cite{ref:ludwig85} and his school.
	It is called {\it the preparation-registration arrow of time}~\cite{ref:harshman98}:
	"A state must be prepared first by a preparation apparatus before an observable can be detected in it by the registration apparatus."
	On applying this causality condition, we notice a serious limitation due to smallness of a quantum system as stated by Dirac~\cite{ref:dirac58}: ``Causality applies only to a system which is left undisturbed.''
	We shall here accept this limitation as a phenomenological principle, in place of the ``collapse of the wave function'' axiom in theories of measurement.

	Starting with the preparation-registration arrow of time and using the "smallness" of quantum system, we will argue in Sec.~\ref{sec:semigroup} that the time evolution of a quantum state or an observable are restricted to $0\leq t <\infty$.
	This type of time evolution is called the semigroup time evolution.
	It will be shown in Sec.~\ref{sec:tabc} that the Hilbert space axiom for the dynamical equations does not accommodate the semigroup time evolution, thus one needs a different boundary condition.
	Hence we introduce the time asymmetric boundary condition which has been obtained before from the requirement that one obtains a unified theory of resonance and decay~\cite{ref:bohm78, ref:bohm97}.
	This boundary condition is provided by a pair of Hardy rigged Hilbert spaces. 
%(The original purpose for using it was to unify the theoretical description of resonance scattering and exponential decay, and the semigroup time evolution was obtained as a by-product~\cite{ref:bohm78}.)
% and to provide Dirac's bra-ket formalism a mathematical counterpart.
	It is this abstract vector space that yields the dispersion relation of the energy wave function of state and of observable as Hardy functions.
%	It should be noted that the Hardy function analytic in the upper complex semi-plane is indeed what has been known as {\it causal transform}~\cite{ref:toll56, ref:nussenzveig72}.
%	We finally define the S-matrix elements and discuss in Sec.~\ref{sec:conclusion} a possible connection between our result and other dispersion relations previously obtained in quantum physics.

%------------- Things to include --------------------------------------
%RHS provides Dirac's bra-ket formalism a mathematical counter-part.
%%%%%%%%%%%%%%%%%%%%%%%%%%%%%%%%%%%%%%%%%%%%%%%%%%%%%%%%%%%%%
\section{Semigroup time evolution due to causality~\label{sec:semigroup}}
%%%%%%%%%%%%%%%%%%%%%%%%%%%%%%%%%%%%%%%%%%%%%%%%%%%%%%%%%%%%%
	We begin our discussion by considering a microsystem that undergoes non-trivial time evolution.
	After reviewing time evolution of states and observables in quantum physics in Sec.~\ref{sec:prepreg}, we formulate a causality condition in Sec.~\ref{sec:causality}.

%--------------------------------------------------------------------
\subsection{Preparation and registration of microsystems -- states and observables~\label{sec:prepreg}}
\begin{figure}[htbp]
\begin{center}
\includegraphics*[height=2.0cm]{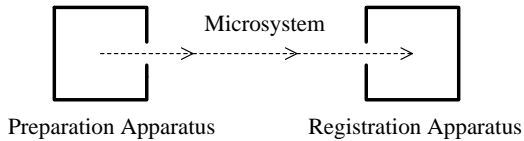}
%figuresize=46x13
\end{center}
\caption{Experiment with microsystems. Microsystems are first prepared by a preparation apparatus then observed by a registration apparatus.
}
\label{fig:prepreg}
\end{figure}

%--------------------------------------------------------------------
	A microsystem is a physical object that can be detected by a macroscopic measuring apparatus~\cite{ref:ludwig85}.
	It is a ``small'' system, e.g., a particle such as electron, in the sense that disturbance due to an observation cannot be neglected\footnote{Dirac has discussed a limit to the finiteness of one's power of observation and the ``smallness'' of the accompanying disturbance~\cite{ref:dirac58}.
	If the disturbance is not negligible, the system being observed is said to be ``small''.}.
	In experiments, as depicted in Fig.~\ref{fig:prepreg}, microsystems are first prepared by a preparation apparatus (e.g., accelerator) and are then subjected to a registration apparatus (e.g., detector).
	A result of experiment, that is a relation between preparation and registration, is described by quantum mechanics.
%%%

	Performing preparation under the same condition would result in an {\it ensemble} of microsystems.
	This ensemble is most generally represented as a quantum state which is described by a density operator $\rho$.
	For the sake of simplicity we restrict ourself here to a special case of it, a {\it pure state}, whose density operator $\rho$ is a projection operator $\rho=\ketbra{\phi}{\phi}$ satisfying $\rho^2 = \rho$.
	The vector $\ket{\phi}$ is a state vector which belongs to an abstract vector space $\Phi_{\rm state}$; in axiomatic quantum mechanics~\cite{ref:neumann55, ref:wightman00}, the Hilbert space $\mathsf{H}$ is commonly chosen for $\Phi_{\rm state}$. 
	Hereafter, we call such a state described by a vector $\ket{\phi}$ the state $\phi$.

	A physical quantity measured with a registration apparatus, like a detector, is described by an observable $\mathcal{O}$.
%	On the other hand, a registration apparatus registers an observable $\mathcal{O}$.
	In the Hilbert space, $\mathcal{O}$ is represented by a selfadjoint operator;  
	its eigenkets form a complete set $\{\ket{o_i}\}$.
	If one performs a single measurement with $\mathcal{O}$ on a microsystem prepared in the state $\phi$, the result one obtains is one of the eigenvalues $o_i$ of $\mathcal{O}$ with a probability $w_i=\abs{\braket{o_i}{\phi}}^2$.
	Hence the expectation value of the measurement is given by $\expect{\mathcal{O}} \equiv \bra{\phi}\mathcal{O}\ket{\phi}=\sum_i o_i\, w_i$.

	In an experimental arrangement like the one depicted in Fig~\ref{fig:prepreg}, every microsystem must first be prepared before it can be registered~\cite{ref:ludwig85}; 
	this statement is called the {\it preparation-registration arrow of time}~\cite{ref:harshman98}.
	According to this principle of causality, the registration of $\mathcal{O}$ can take place anytime but only after preparation of the state $\phi$ has been completed.

	The time evolution of the state is described by a dynamical equation.
	In the Schr\"odinger picture the state evolves in time obeying the Schr\"odinger equation\footnote{We take natural unit system in which $\hbar=1$.},
\begin{subequations}
\begin{align}
i\frac{d}{dt}\ket{\phi(t)} = H\,\ket{\phi(t)}
\label{eq:schrodingereq}
,
\intertext{and observables are kept fixed in time. Here $H$ is the Hamiltonian operator of the microsystem under consideration. 
	Equivalently in the Heisenberg picture the observable evolves in time obeying the Heisenberg equation,}
-i \frac{d\mathcal{O}(t)}{dt} = \left[H, \mathcal{O}(t) \right]
\label{eq:heisenbergeq}
,
\end{align}
\label{eq:dynamicaleqs}
\end{subequations}
and the state vector $\ket{\phi}$ is time independent.
	If $\ket{\phi}$ is prepared at time $t_0$, the solutions of Eqs.~\eqref{eq:schrodingereq} is given by,
\begin{subequations}
\begin{align}
\ket{\phi(t)} &= e^{-iH(t-t_0)}\,\ket{\phi}
\label{eq:solofstate}
,
\intertext{provided $\frac{d}{dt}H=0$, and the solution of Eq.~\eqref{eq:heisenbergeq} is given by}
\mathcal{O}(t) &= e^{iH(t-t_0)}\,\mathcal{O} e^{-iH(t-t_0)}
\label{eq:heisenbergoperator1}
.
\end{align}
\label{eq:solutions}
\end{subequations}
	The expectation value for a measurement of an observable $\mathcal{O}$ in the state $\phi$ is expressed as
\begin{subequations}
\begin{align}
\expect{\mathcal{O}}_t 
&= \bra{\phi(t)}\mathcal{O}\ket{\phi(t)}\ \mbox{in the Schr\"odinger picture}\\
&= \ \ \bra{\phi}\mathcal{O}(t)\ket{\phi} \ \ \ \mbox{in the Heisenberg picture}
.
\end{align}
\label{eq:timedepexpect}
\end{subequations}

	Note that a {\it boundary condition} has been tacitly assumed on integrating the Schr\"odinger equation~\eqref{eq:schrodingereq} to obtain the solution~\eqref{eq:solofstate}.
	One solves the dynamical Eq.~\eqref{eq:schrodingereq} for the boundary condition that the solutions should be elements of the Hilbert space $\mathsf{H}$, i.e., $\Phi_{\rm{state}} = \mathsf{H}$.
	Then as a consequence of the Stone-von neumann theorem, the time parameter $t$ in Eqs.~\eqref{eq:solutions}~and~\eqref{eq:timedepexpect} must range over the whole real line, $-\infty<t<\infty$~\cite{ref:stone-vonneumann}.
	In the following section, we will show that this does not meet the physical requirements of causality. 

%	In the next section, we discuss a physically allowed range of the time parameter $t$.

%------------------------------------------------------------------------------------
\subsection{The ensemble of beginnings of time and the limitation on time evolution~\label{sec:causality}}
%------------------------------------------------------------------------------------
	The preparation-registration arrow of time we shall discuss now is a general statement about ``cause and effect'' whose application is not necessarily restricted to a microsystem; it could equally be applied to any ``large'' object which obeys the laws of classical physics.
	In quantum physics, however, there is a limitation on application of causality due to the ``smallness'' of a microsystem.
	This principle given by Dirac\footnote{See section 1 of Ref~\cite{ref:dirac58}.} we now quote:
	``Causality applies only to a system which is left undisturbed. If a system is small, we cannot observe it without producing a serious disturbance and hence we cannot expect to find any causal connexion between the results of our observations.''
	As a consequence, the best we can expect for the observations (measurements) are probability predictions.
	In accordance with these phenomenological consideration, we conclude that {\it a description of a causal connection by the dynamical equations is possible only between two successive measurements, the preparation of the state and the registration of the observable, during which the microsystem is left undisturbed\footnote{See section 27 of Ref~\cite{ref:dirac58}.} .}
	Once a microsystem is prepared in a state, any previous information about this microsystem is destroyed, and hence it is impossible to trace the system back with the dynamical equations to a time before the state was prepared.
%the preparation of a microsystem completely destroys any previous information about the microsystem, and it is unable to trace back with the dynamical equations.

\begin{figure}[htbp]
\begin{center}
\includegraphics*[height=4cm]{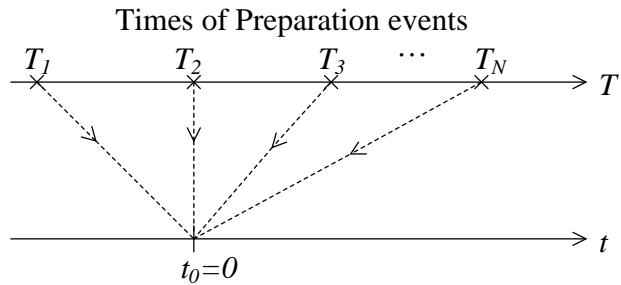}
%figuresize=46x13
\end{center}
\caption{The ensemble of beginnings of time for the state $\phi$.}
\label{fig:ebot}
\end{figure}

	To express this phenomenological principle in a mathematical form, let us first discuss the preparation of microsystems.	
%	As we have emphasized, a state vector represents an ensemble of microsystems.
	For definiteness, we here consider an ensemble of $N$ microsystems (e.g., electrons) all to be prepared in the same state $\phi$.
	(Here $N$ can be arbitrary large number.)
%	Since identical experiments can be carried out at anytime in the future and as well anytime in the past\footnote{but certainly not before the beginning of universe}
	In a laboratory, an experimenter would complete this preparation by performing measurements under the same condition, and the preparation events would be recorded referring to the experimenter's clock $T$.
	Let us denote the preparation instant of the $i$-th microsystem by $T_i$.
%\footnote{Since identical experiments can be carried out at anytime in the future or in the past, $T$ can take any real value, but certainly not before the beginning of universe.}.
	As a result of the preparation, one obtains an ensemble of preparation events at $\{T_1, T_2, \cdots, T_N\}$, that we refer to {\it the ensemble of beginnings of time}~\cite{ref:bohm06}.
	Since this ensemble describes the times at which the same state $\phi$ is prepared, all of these times must be mapped onto exactly the same point $t=t_0$ as shown in Fig.~\ref{fig:ebot}.
%%%
%%%%%	
	Without loss of generality one can choose $t_0$ to be $0$, and thus we denote the ensemble of beginnings of time as the time
\begin{align}
t=0\ : \ \{T_1, T_2, \cdots, T_N\}\ \mbox{for the state $\phi$}
.
\label{eq:begoftime}
\end{align}

	Note that each $T_i$ in this ensemble is completely individual, i.e., there is no correlation.
	In order to illustrate this fact, we shall here present two completely different ways of preparing the same state.
% but completely equivalent ways of the preparation.	
%	At one extreme, one can prepare all of the $N$ microsystems simultaneously in the state $\phi$.
	At one extreme, one can prepare all of the $N$ microsystems simultaneously (but in general at different places) at the same time $T^0$ on the experimenter's clock.
	In this way all of the preparation times in Eq.~\eqref{eq:begoftime} would be the same on the experimenters clock but the $N$ individual microsystems may be at different location but otherwise the same.
	Then
\begin{align}
T^0=T_1=T_2= \cdots = T_N
\label{eq:begoftimesc}
.
\end{align}
%
%	Although this is possible in principle it is very difficult to achieve in practice.
	At the other extreme, one can work with only one single microsystem and repeat the preparation to the state $\phi$ for $N$ times.
	Examples of this latter case are the single-ion experiments~\cite{ref:ionexp}, where the microsystem (a single ion) is prepared in an unstable state at $N=100-200$ times.
% and for $N=150$ times in Ref.~\cite{ref:5d}.
	In such an experiment, preparation events take place at {\it different} times as 
\begin{align}
T_1<T_2<\cdots <T_N
\label{eq:differentbeginnings}
,
\end{align}
but again they are all mapped onto $t=0$.

\begin{figure}[htbp]
\begin{center}
\includegraphics*[height=4cm]{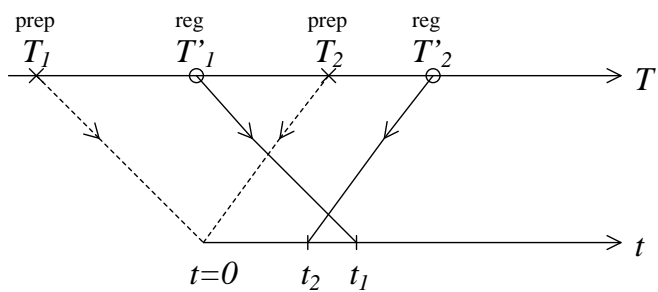}
%size=49x23
\end{center}
\caption{
The preparation times ($T_1$ and $T_2$) and the registration times ($T_1'$ and $T'_2$) in the experimenter's time and the quantum mechanical parameter ($t_1$ and $t_2$).
All of the preparation times (beginnings of time) are mapped onto $t=0$.
Then the time intervals between the preparation and the registration are mapped onto the time parameter $t$.
}
\label{fig:time}
\end{figure}

	Now we turn our attention to the time evolution of the states and observables.
	For each of the $N$ preparation events~\eqref{eq:begoftime}, there corresponds a registration event of observable $\mathcal{O}$.
	Let us here denote the time of registration event (e.g., the time of decay in the case of the single ion experiment) by $T'_i$ for the $i$-th individual microsystem.
	Then one obtains the ensemble of the ``ends'' of time,
\begin{align}
\left\{T'_1, T'_2, \cdots , T'_N \right\} \ \mbox{for observable $\mathcal{O}$}
.
\label{eq:endoftime}
\end{align}
	In accordance with the preparation-registration arrow of time, the registration event at $T'_i$ must always be later than its corresponding preparation event at $T_i$, that is 
\begin{align}
T'_i \geq T_i \ \ \mbox{for $i=1,2,\cdots,N$}
\label{eq:prepreg-math}
\end{align}
where equality holds when the registration follows immediately after the preparation.
	Finally, as shown in Fig~\ref{fig:time}, from the ensembles~\eqref{eq:begoftime}~and~\eqref{eq:endoftime} one obtains an {\it ensemble} of time intervals $\{t_1, t_2, \cdots, t_N\}$ during which the microsystems were left undisturbed from any measurements,
\begin{align}
t_1 &= T'_1-T_1
,
\nonumber
\\
t_2 &= T'_2-T_2
,
\nonumber
\\
&\ \ \vdots
\nonumber
\\
t_N &= T'_N-T_N
.
%\nonumber
\label{eq:timeparam}
\end{align}
%
%%%
%%%%%
	Here $t_i\geq 0$ must hold by the preparation-registration arrow of time~\eqref{eq:prepreg-math}.
	Now we draw a conclusion from Eq.~\eqref{eq:timeparam}: {\it the time parameter $t$ represents the ensemble of time-intervals during which the system evolved undisturbed,
\begin{align}
t\ : \ \{t_1, t_2, \cdots, t_N \}
\label{eq:ensembleoftime}
.
\end{align}
	Because every one of the time intervals is always positive we have}
\begin{align}
0\leq t < \infty
\label{eq:timeasymmetry}
.
\end{align}
	This is a general expression of the preparation-registration arrow of time in quantum physics.

	In accordance with Eq.~\eqref{eq:timeasymmetry}, the expectation value~\eqref{eq:timedepexpect} makes physical sense only for $t\geq0$.
	This means that the boundary condition for the Schr\"odinger equation~\eqref{eq:schrodingereq} must be chosen such that its general solution is given by
\begin{subequations}
\begin{align}
\ket{\phi(t)} &= e^{-iHt}\,\ket{\phi} \quad\mbox{for $0\leq t<\infty$ only}
\label{eq:causalstate}
.
\intertext{Alternatively in the Heisenberg picture, the solution of Eq.~\eqref{eq:heisenbergeq} must be given by}
\mathcal{O}(t)&= e^{iHt}\,\mathcal{O}\,e^{-iHt} \quad \mbox{for $0\leq t<\infty$ only}
.
\label{eq:causalobs}
\end{align}
\label{eq:causaltimeevolution}
\end{subequations}
	Such a time evolution~\eqref{eq:causaltimeevolution} is called {\it semigroup}, because the time evolution operator $e^{-iHt}$ of the state vector has no inverse $(e^{-iHt})^{-1}$.

%%%%%%%%%%%%%%%%%%%%%%%%%%%%%%%%%%%%%%%%%%%%%%%%%%%%%%%%%%%%%%%%%%%%%%%%%%%%%%%%
\section{The Boundary condition for the semigroup time evolution\label{sec:tabc}}
%%%%%%%%%%%%%%%%%%%%%%%%%%%%%%%%%%%%%%%%%%%%%%%%%%%%%%%%%%%%%%%%%%%%%%%%%%%%%%%%

	The solutions of the Schr\"odinger equation depend upon boundary condition, i.e., the conditions that which space the solutions $\ket{\phi(t)}$ of the dynamical equation~\eqref{eq:schrodingereq} must belong to.
	In order to obtain Eq.~\eqref{eq:causaltimeevolution}, one has to use a boundary condition which leads to a semigroup time evolution.

%-------------------------------------------------------------------------------
\subsection{Why not the Hilbert space\label{sec:hilbert}}
%-------------------------------------------------------------------------------
	Under the Hilbert space boundary condition (or the Hilbert space axiom of the standard (axiomatic) quantum mechanics), the state vectors are postulated to be elements of the Hilbert space, $\ket{\phi(t)}\in\mathsf{H}$.
	As a consequence of this, one obtains from the Stone-von Neumann theorem~\cite{ref:stone-vonneumann} that the solutions of the differential equation~\eqref{eq:schrodingereq} with a selfadjoint Hamiltonian $H$ are given by unitary {\it group} $U^\dagger(t)\equiv e^{-iHt}$ with $-\infty<t<\infty$, which means the time evolution in Eq.~\eqref{eq:solofstate} is given by
\begin{align}
\ket{\phi(t)} = e^{-iHt}\,\ket{\phi}\quad \mbox{with $-\infty<t<\infty$}
\label{eq:unitarygroup}
.
\end{align}
	This disagrees with our phenomenological conclusion~\eqref{eq:timeasymmetry}, which suggested that Eq.~\eqref{eq:causalstate} must hold.

	In scattering theory, one overcomes the discrepancy between~\eqref{eq:unitarygroup} and the phenomenological result~\eqref{eq:causalstate}  by employing the propagator or retarded Green's function~\cite{ref:goldberger64, ref:newton82} $G(t)$,
\begin{align}
G(t) = \theta(t)\ e^{-iHt} 
= \left\{ 
\begin{array}{ll} 
e^{-iHt} & \mbox{for $0\leq t<\infty$}, \\ 0 & \mbox{for $-\infty<t<0$}
.
\end{array} 
\right.
\label{eq:propagator}
\end{align}
	This removes the unwanted negative-time part, however, such a solution of Eq.~\eqref{eq:schrodingereq} no longer belongs to the Hilbert space.
	The impossibility of Hilbert space vectors $\ket{\phi(t)}$ with the property that
\begin{align}
\int_{-\infty}^\infty dt\ \bra{\phi(t)}\mathcal{O}\ket{\phi(t)} = 0
,
\end{align}
as would be the case for $\ket{\phi(t)}=G(t)\ket{\phi}$ given by Eq.~\eqref{eq:propagator}, is also the consequence of a mathematical theorem~\cite{ref:hegerfeldt94}.

	Thus, although the Hilbert space has been a successful choice for eigenstates of discrete energy which have trivial time evolution (stationary states), it does not accommodate the semigroup time evolution~\eqref{eq:causaltimeevolution} derived from the phenomenological causality condition we employed.

%----------------------------------------------------------
\subsection{The Hardy space for states}
%----------------------------------------------------------
	Now we show that one can formulate {\it the time asymmetric boundary condition}~\cite{ref:bohm97} for the time symmetric differential equation~\eqref{eq:schrodingereq} and that it leads to the semigroup time evolution.	
	Since the formulation of time asymmetric boundary conditions requires the rigged Hilbert space (RHS)~\cite{ref:gelfand64}, we shall here present some facts and notations necessary to include Dirac's bra-and-ket formalism the mathematical theory provided by the Schwartz spaces and a a generalization thereof.

	According to Dirac~\cite{ref:dirac58} every state vector $\ket{\phi}$ is expanded with respect to the eigenkets of an operator with a continuous as well as discrete spectrum.
	In order for us to work with the simplest representation of the time evolution operator $e^{-iHt}$, we here employ the energy and angular momentum eigenkets $\ket{E\,\ell\,\ell_3}$ that satisfy the eigenvalue equations,
\begin{subequations}
\begin{align}
H\,\ket{E\,\ell\,\ell_3} &= E\, \ket{E\,\ell\,\ell_3},\quad 0<E<\infty
\label{eq:diracket1}
\\
\bm{L}^2 \,\ket{E\,\ell\,\ell_3} &= \ell(\ell+1)\,\ket{E\,\ell\,\ell_3},\quad \ell=0,1,2,\cdots
\\
L_3\,\ket{E\,\ell\,\ell_3} &= \ell_3\,\ket{E\,\ell\,\ell_3},\quad \ell_3=-\ell, -\ell+1, \cdots, \ell
,
\end{align}
\label{eq:energyandangularcont}
\end{subequations}
and a normalization
\begin{align}
\braket{E'\ell'\ell_3'}{E\,\ell\,\ell_3}=\delta(E'-E)\,\delta_{\ell'\ell}\,\delta_{\ell'_3 \ell_3}
\label{eq:deltanorm}
.
\end{align}
	Assuming a spherically symmetric Hamiltonian, $[H,\bm{L}]=0$, and neglecting the spin for simplicity, the expansion of the state vector~\cite{ref:dirac58} is given by
\begin{align}
\ket{\phi} = \sum_{\ell\ell_3} \int_0^\infty dE\ \ket{E\,\ell\,\ell_3}\,\phi_{\ell\ell_3}(E)
\label{eq:generalexpansion}
,
\end{align}
where $\phi_{\ell\ell_3}(E)\equiv\braket{E\,\ell\,\ell_3}{\phi}$ is the energy wave function\footnote{This is more commonly called the partial wave, but to make a clear distinction between partial wave of S-matrix (scattering amplitude) and this wave function we shall consistently use a terminology ``energy wave function'' for the representative.}.	
	This expansion provides an one-to-one correspondence between vector $\ket{\phi} \in \Phi_{\rm state}$ for solutions of the Schr\"odinger equation and {\it the space for the energy wave functions} $\phi_{\ell\ell_3}(E)$.
	For example, if one choses for $\phi_{\ell\ell_3}(E)$ smooth and rapidly decreasing functions of $E$, then their space is the Schwartz space $\mathcal{S}$~\cite{ref:wightman00}, i.e., $\phi_{\ell\ell_3}(E)\in\mathcal{S}(\mathbb{R}_+)$, where $\mathbb{R}_+$ denotes the positive real energy axis.
	To every function $\phi_{\ell\ell_3}(E)\in\mathcal{S}(\mathbb{R}_+)$, there corresponds an abstract vector $\ket{\phi}$ given by~\eqref{eq:generalexpansion}.
	To the set of all $\left\{ \phi_{\ell\ell_3}(E) \right\}=\mathcal{S}(\mathbb{R}_+)$ there corresponds a dense subspace $\Phi$ of the Hilbert space, $\Phi\subset\mathsf{H}$.
	The state vector $\ket{\phi}$ is then an element of the abstract vector space, $\ket{\phi}\in\Phi$.
	On the other hand, the eigenkets~\eqref{eq:energyandangularcont} with normalization~\eqref{eq:deltanorm} are not elements of $\Phi$ or $\mathsf{H}$; 
	they are elements of the abstract vector space $\Phi^\times$ of linear continuous functionals $\mathcal{S}^\times$ (distributions, such as the delta function) over the Schwartz space~\cite{ref:wightman00}.
	This means that $\ket{E\,\ell\,\ell_3}\in\Phi^\times$ where $\Phi^\times$ denotes the space of all antilinear continuous functionals on $\Phi$.
	These three abstract vector spaces together make up a rigged Hilbert space called the Schwartz RHS\footnote{Dirac has emphasized in Ref~\cite{ref:dirac58} ``The bra and ket vectors that we now use form a more general space than a Hilbert space.'' The functional space $\Phi^\times$ is this general space. It follows from this that the operators in Eq.~\eqref{eq:energyandangularcont} are more general than those in the Hilbert space, and the precise definition of~\eqref{eq:diracket1} is $$\braket{H\phi}{E\,\ell\,\ell_3}\equiv\bra{\phi}H^\times\ket{E\,\ell\ell_3}=E\braket{\phi}{E\,\ell\,\ell_3}$$ for all $\ket{\phi}\in\Phi$, where $H^\times$ is the unique extension of the operator $H^\dagger$ from the Hilbert space $\mathsf{H}=\mathsf{H}^\times$ to $\Phi^\times$.},
\begin{align}
\Phi\subset\mathsf{H}\subset\Phi^\times
.
\end{align}

	In order to formulate time asymmetric boundary conditions for the solutions of the dynamical equations~\eqref{eq:dynamicaleqs}, one has to go one step further:
	In the expansion~\eqref{eq:generalexpansion}, require that the energy wave function $\phi_{\ell\ell_3}(E)$ not only be a Schwartz function $\mathcal{S}(\mathbb{R}_+)$ but also be {\it analytically continued in the lower-half complex (energy) semi-plane} and satisfies the following condition:
\begin{align}
\int_{-\infty}^\infty dE\, \abs{\phi^+_{\ell\ell_3}(E+iy)}^2 < \infty \ \mbox{for any $y<0$}
\label{eq:hardycriteria1}
.
\end{align}
	Here the superscript ``$+$'' indicates the analyticity of the wave function\footnote{This ``$+$'' sign has its origin in the scattering theory; in-state prepared by an accelerator has been conventionally denoted by $\ket{\phi^+}$.}.
	Such a function is called a {\it Hardy function analytic in the lower complex semi-plane}, or {\it Hardy function from below} in short~\cite{ref:bohm97, ref:hardyfunction}, and we denote its function space\footnote{This is a function space for ``smooth'' Hardy functions. The function space $\mathcal{S}_-(\mathbb{R}_+)$ is defined by the intersection of the Schwartz space $\mathcal{S}$ and the function space of the Lebesgue square-integrable functions of Hardy class $\mathcal{H}^2_-$ restricted to $\mathbb{R}_+$, i.e.,  $\mathcal{S}_-(\mathbb{R}_+)\equiv\left[\mathcal{S}\cap\mathcal{H}_-^2 \right]_{\mathbb{R}_+}$. This allows one to obtain the Hardy RHS.} by $\mathcal{S}_-(\mathbb{R}_+)$,
\begin{align}
\phi^+_{\ell\ell_3}(E)\in\mathcal{S}_-(\mathbb{R}_+)
.
\end{align}
	Note that in Eq.~\eqref{eq:hardycriteria1} it seems problematic that the function values on the negative real-axis (of energy), which cannot be reached by experiment, are needed to perform the integral. 
	For Hardy functions these values can be reconstructed from its boundary values on the positive-real axis\footnote{This is by virtue of the van Winter theorem of Hardy functions}~\cite{ref:vanwinter}, and hence the Hardy function $\phi^+_{\ell\ell_3}(E)$ is completely determined by their values for the physical energies $0<E<\infty$.
	It may also be noted that Eq.~\eqref{eq:hardycriteria1} (along with the required analyticity) is necessary and sufficient condition for that the energy wave function fulfills the Hilbert transform,
\begin{subequations}
\begin{align}
{\rm Re}\ \phi_{\ell\ell_3}^+(E) &=  -\frac{1}{\pi}\ {\rm P} \int_{-\infty}^\infty d\omega\ \frac{{\rm Im}\ \phi_{\ell\ell_3}^+(\omega)}{\omega-E}\ ,
\\
{\rm Im}\ \phi_{\ell\ell_3}^+(E) &=  \ \ \frac{1}{\pi}\ {\rm P} \int_{-\infty}^\infty d\omega\ \frac{{\rm Re}\ \phi_{\ell\ell_3}^+(\omega)}{\omega-E}\ ,
\end{align}
\label{eq:statedisp}
\end{subequations}
where ${\rm P}$ designates the Cauchy principal value.
	In physics, Eq.~\eqref{eq:statedisp} is called {\it dispersion relation}~\cite{ref:goldberger64, ref:jackson98} which often appears in connection with causality.
%\footnote{Eq.~\eqref{eq:statedisp} differs from the dispersion theory of S-matrix in which T-matrix is required to satisfy the (subtracted) dispersion relations.}.

	It has been shown by Gadella that the Hardy functions provide a rigged Hilbert space, called the {\it Hardy RHS}, denoted by $\Phi_-\subset\mathsf{H}\subset\Phi_-^\times$~\cite{ref:gadella89}.
	Stated generally, the time asymmetric boundary condition requires the state $\phi$ to be described by a state vector $\ket{\phi^+}$ in the Hardy space $\Phi_-$.
	In the Hardy RHS, there exists for every vector $\ket{\phi^+}\in\Phi_-$ a set of eigenkets $\ket{E\,\ell\,\ell_3\,^+}\in\Phi^\times_-$  \footnote{More precisely, the Hamiltonian of Eq.~\eqref{eq:hardyket1} is to be denoted by $H_-^\times$, as the Hamiltonian operators $H_\pm$ on $\Phi_\pm$ are defined as the restriction of the selfadjoint Hamiltonian on the Hilbert space, $\overline{H}$, to the dense subspaces $\Phi_\pm$ of $\mathsf{H}$. The operator $H_\pm^\times$ are the uniquely defined extensions of the operator $\overline{H}=H^\dagger$ to the spaces $\Phi_\pm^\times$. In this article, however, we denote all the different Hamiltonian operators as $H$, for their mathematical differences are immaterial for our purpose.}
\begin{subequations}
\begin{align}
H\,\ket{E\,\ell\,\ell_3\,^+} &= E\, \ket{E\,\ell\,\ell_3\,^+},\quad 0<E<\infty
\label{eq:hardyket1}
\\
\bm{L}^2 \,\ket{E\,\ell\,\ell_3\,^+} &= \ell(\ell+1)\,\ket{E\,\ell\,\ell_3\,^+},\quad \ell=0,1,2,\cdots
\\
L_3\,\ket{E\,\ell\,\ell_3\,^+} &= \ell_3\,\ket{E\,\ell\,\ell_3\,^+},\quad \ell_3=-\ell, -\ell+1, \cdots, \ell
,
\end{align}
\label{eq:hardyket+}
\end{subequations}
with a normalization
\begin{align}
\braket{^+E'\ell'\ell_3'}{E\,\ell\,\ell_3\,^+}=\delta(E'-E)\,\delta_{\ell'\ell}\,\delta_{\ell'_3 \ell_3}
\label{eq:deltanormstate}
,
\end{align}
such that $\ket{\phi^+}$ can be written as
\begin{align}
\ket{\phi^+} = \sum_{\ell\ell_3} \int_0^\infty dE\ \ket{E\,\ell\,\ell_3\,^+}\,\phi^+_{\ell\ell_3}(E)
\label{eq:statepartialwave+}
,
\end{align}
where $\phi^+_{\ell\ell_3}(E) \equiv \braket{^+E\,\ell\,\ell_3}{\phi^+}$.
	This means that one has the one-to-one correspondence between the state vector $\ket{\phi^+}$ and its energy wave function $\phi^+_{\ell\ell_3}(E)$ as
\begin{align}
\ket{\phi^+} \in \Phi_- \longleftrightarrow \phi^+_{\ell\ell_3}(E) \in \mathcal{S}_-(\mathbb{R}_+)
\label{eq:hardybc2}
.
\end{align}

	We shall now show that the property of the Hardy functions~\eqref{eq:hardycriteria1} leads to the semigroup time evolution.
	For the state $\phi$ prepared as $\ket{\phi^+}$ at $t=0$, its time evolved state vector is given by $\ket{\phi^+(t)}=e^{-iHt}\ket{\phi^+}$.
	The basisket expansion~\eqref{eq:statepartialwave+} of this vector is then given by 
\begin{align}
\ket{\phi^+(t)} 
%= \int_0^\infty dE\ e^{-iHt} \ket{E\,\ell\,\ell_3\,^+}\,\phi_{\ell\ell_3}^+(E)
&= \sum_{\ell\ell_3}\int_0^\infty dE\ \ketbra{E\,\ell\,\ell_3\,^+}{^+E\,\ell\,\ell_3}e^{-iHt}\ket{\phi^+}
\nonumber \\
&= \sum_{\ell\ell_3}\int_0^\infty dE\ \ket{E\,\ell\,\ell_3\,^+}\left[e^{-iEt}\phi_{\ell\ell_3}^+(E)\right]
\label{eq:calc1}
,
\end{align}
where Eq.~\eqref{eq:hardyket1} has been used to obtain the last expression. 	
	In order that the time evolved state $\ket{\phi^+(t)}$ fulfills the time asymmetric boundary condition, that is in order that $\ket{\phi^+(t)}\in\Phi_-$, its time-dependent energy wave function $e^{-iEt}\phi^+_{\ell\ell_3}(E)$ given in Eq.~\eqref{eq:calc1} must satisfy Eq.~\eqref{eq:hardycriteria1},
\begin{align}
\int_{-\infty}^\infty dE\ \abs{e^{-i(E+iy)t}\phi_{\ell\ell_3}^+(E+iy)}^2
=\int_{-\infty}^\infty dE\ e^{2ty}\abs{\phi_{\ell\ell_3}^+(E+iy)}^2
<\infty
\label{eq:calc2}
.
\end{align}
	For this integral to converge for arbitrary large negative $y$, the time parameter $t$ cannot be negative, so that only $0\leq t <\infty$ is allowed.
	Hence we obtain the time evolution of the state
\begin{align}
\ket{\phi^+(t)} &= e^{-iHt}\,\ket{\phi^+} \quad\mbox{for $0\leq t<\infty$ only}
\label{eq:causalstate+}
,
\end{align}
if we restrict ourself to the Hardy space $\Phi_-$.
	This is exactly the desired semigroup time evolution~\eqref{eq:causalstate}.

	In practice, the square-modulus of the wave function $\abs{\phi_{\ell\ell_3}^+(E)}^2$ is interpreted as the energy-angular distribution of the prepared microsystems, which is the characteristic of preparation apparatus.
	For example, if the state $\phi$ is prepared in such a way that its energy wave function be a Hardy function from below,
\begin{align}
\phi_{\ell\ell_3}^+(E) = \frac{C_{\ell\ell_3}}{E-(a + ib/2)}
\ \in\mathcal{S}_-(\mathbb{R}_+)
,
\end{align}
where the expansion coefficient $C_{\ell\ell_3}$ is complex in general and $a,\,b>0$, then its energy distribution function $f(E)$ is a Lorentzian function,
\begin{align}
f(E)\equiv\sum_{\ell\ell_3}\abs{\phi_{\ell\ell_3}^+(E)}^2 = \frac{\sum_{\ell\ell_3}\abs{C_{\ell\ell_3}}^2}{(E-a)^2 + (b/2)^2}
\label{eq:statelorentzian}
,
\end{align}
in which the peak energy is given by $a$ and the FWHM by $b$.
	The coefficients $\abs{C_{\ell\ell_3}}^2$ describe the angular distribution that satisfy the normalization condition
\begin{align}
||\phi^+||^2 &= \braket{\phi^+}{\phi^+} = \int_0^\infty dE\, f(E) = 1
%||\phi^+||^2 &= \braket{\phi^+}{\phi^+} = \sum_{\ell\ell_3} \int_0^\infty dE\, \abs{\phi_{\ell\ell_3}^+(E)}^2 = 1
%			  = \frac{\pi + 2\arctan{(a/b)}}{2b} \sum_{\ell\ell_3}\abs{C_{\ell\ell_3}}^2  = 1,
.
\end{align}
%  
%where the definition~\eqref{eq:statelorentzian} has been used.

%-------------------------------------------------------------------------
\subsection{Transition probability and the Hardy space for observables}	
%-------------------------------------------------------------------------
	We here concern with a selective measurement~\cite{ref:haag96, ref:sakurai94, ref:schwinger59} (or filteration) in which a registration apparatus is designed to select only one of the eigenvectors of $\mathcal{O}$, or a particular linear combination of them.
	The observable that represents the measurement is given by
\begin{align}
\mathcal{O}_\psi = \ketbra{\psi}{\psi}
\label{eq:projectionop}
,
\end{align}
where $\ket{\psi}$ is a normalized vector which can be expanded as
\begin{align}
\ket{\psi}=\sum_{\ell\ell_3} \int_0^\infty dE\, \ket{E\,\ell\,\ell_3} \psi_{\ell\ell_3}(E)
\label{eq:obsexp}
.
\end{align}
	Here $\psi_{\ell\ell_3}(E)\equiv\braket{E\,\ell\,\ell_3}{\psi}$ is a wave function characterizing the measurement\footnote{For example, if the measurement is to select $\ell=0$ eigenvector, then $\psi_{\ell\ell_3}(E)$ is nonzero only if $\ell=\ell_3=0$. For an observable having continuous eigenvalues, such as a Hamiltonian $H$, one can make an ``almost eigenket''. In the case of Eq.~\eqref{eq:obsexp}, it is given by a vector $\ket{\psi}$ in which $\abs{\psi_{\ell\ell_3}(E)}^2$ has a very sharp peak with a finite width~\cite{ref:bohm01book}.}.
	The observable~\eqref{eq:projectionop} satisfies $\mathcal{O}_\psi^2=\mathcal{O}_\psi$; the eigenvalue of $\mathcal{O}_\psi$ is $1$ (`affirmative') or $0$ (`negative') only, with $\ket{\psi}$ being the eigenvector belonging to the eigenvalue $1$ and all vectors orthogonal to $\ket{\psi}$ having the eigenvalue $0$.
	For $\mathcal{O}_\psi$ is completely specified by a vector $\ket{\psi}$, we call Eq.~\eqref{eq:projectionop} the {\it observable} $\psi$ hereafter.
%	Since this observable is a projection operator that projects any vector onto a one-dimensional space spanned by $\ket{\psi}$, we denote this space by $\Psi_{\rm obs}$,  and we call also the vector $\ket{\psi}\in\Psi_{\rm obs}$ itself yes-or-no observable\footnote{More precisely, it is really the ray $\{e^{i\alpha} \ket{\psi}\ |\ \alpha \ \mbox{real} \}$ which corresponds to the operator $\ketbra{\psi}{\psi}$.}.

	The result one obtains for a number of selective measurements is a {\it Born probability}, or more commonly called a {\it transition probability}.
	In the Schr\"odinger picture, by substituting Eq.~\eqref{eq:projectionop} into Eq.~\eqref{eq:timedepexpect}, the transition probability $\mathcal{P}(t)$ to detect the observable $\psi$ in the state $\phi$ is a special case of expectation value given as
\begin{align}
\mathcal{P}(t) \equiv \langle \mathcal{O_\psi} \rangle_t
= \braket{\phi^+(t)}{\psi}\braket{\psi}{\phi^+(t)} 
&= \abs{\braket{\psi}{\phi^+(t)}}^2\ \mbox{for $0\leq t<\infty$}
,
\end{align}
\label{eq:tranprob}
where the time evolved state vector is given by Eq.~\eqref{eq:causalstate+}.
	This quantity certainly is probability as its values are bound to be continuous real between 0 and 1 due to the Cauchy-Schwartz inequality,
\begin{align}
0\leq \abs{\braket{\psi}{\phi^+(t)}}^2 \leq \braket{\psi}{\psi}\, \braket{\phi^+(t)}{\phi^+(t)}=1
,
\end{align}
provided the normalization $\braket{\phi^+(t)}{\phi^+(t)}=\braket{\psi}{\psi}=1$.
	This inequality holds as long as the scalar product (bracket) between $\ket{\psi}$ and $\ket{\phi^+}$ is defined in positive hermitian form~\cite{ref:bohm01book}.
%	This means that in general $\Psi_{\rm state}$ need not be the same as $\Psi_{\rm obs}$ but they can be at most some different subspaces of the same linear space.

	The question of immediate concern is, Is the vector $\ket{\psi}$ in the Hardy space $\Phi_-$?
	To answer it, one has to examine the Heisenberg picture.
	The time evolved observable $\mathcal{O}_\psi(t)$ is given by substituting Eq.~\eqref{eq:projectionop} into Eq.~\eqref{eq:causalobs} as
\begin{align}
\mathcal{O}_\psi(t)=e^{iHt}\ketbra{\psi}{\psi}e^{-iHt} = \ketbra{\psi(t)}{\psi(t)}
,
\end{align}
where we have defined a time evolved vector
\begin{align}
\ket{\psi(t)}\equiv e^{iHt}\ket{\psi}\quad \mbox{for $0\leq t<\infty$}
\label{eq:solofobs}
.
\end{align}
	This expression is similar to Eq.~\eqref{eq:causalstate} but the sign of exponent is opposite.
	This means that if one had taken $\ket{\psi}\in\Phi_-$ the time evolution of $\ket{\psi(t)}$ would be for $-\infty<t\leq0$, which contradicts Eq.~\eqref{eq:solofobs}.
	Thus the vector $\ket{\psi}$ is {\it not} an element of $\Phi_-$. 
	Also, it cannot be an element of the Hilbert space either because it does not lead to the semigroup time evolution~\eqref{eq:solofobs}.

	Thus for the observable $\psi$ one needs yet another boundary condition for solutions of the Heisenberg equation~\eqref{eq:heisenbergeq}.
	As can be seen in Eq.~\eqref{eq:calc2}, the sign of exponent is directly connected to the domain of analyticity of the wave function, namely the lower-half energy plane, for obtaining the semigroup time evolution for $0\leq t<\infty$.
	Our previous discussion about Eq.~\eqref{eq:solofobs} therefore suggests that the wave function $\psi_{\ell\ell_3}(E)$ in the basisket expanstion~\eqref{eq:obsexp} is to be analytic in the {\it uppper}-half energy plane.
	Following this observation, we take the other type of Hardy function, the Hardy function analytic in the {\it upper} complex semi-plane, or Hardy function from {\it above} in short~\cite{ref:hardyfunction}, for this energy wave function\footnote{Here $\mathcal{S}_+(\mathbb{R}_+)\equiv[\mathcal{S}\cap\mathcal{H}^2_+]_{\mathbb{R}_+}$}:
\begin{align}
\psi^-_{\ell\ell_3}(E) \in \mathcal{S}_+(\mathbb{R}_+)
,
\label{eq:obswave}
\end{align}
where the superscript ``$-$'' is to indicate that $\psi^-_{\ell\ell_3}(E)$ is analytic and Hardy in the {\it upper}-half of the complex (energy) plane.
	The energy wave function of observable $\psi$ then satisfies the defining criterion of the Hardy function from above,
\begin{align}
\int_{-\infty}^\infty dE\, \abs{\psi_{\ell\ell_3}^-(E+iy)}^2 < \infty \ \mbox{for any $y>0$}
\label{eq:hardycriteria-}
.
\end{align}
	And in place of the dispersion relations~\eqref{eq:statedisp}, the following relation holds:
\begin{subequations}
\begin{align}
{\rm Re}\ \psi_{\ell\ell_3}^-(E) &=  \ \ \frac{1}{\pi}\ {\rm P} \int_{-\infty}^\infty d\omega\ \frac{{\rm Im}\ \psi_{\ell\ell_3}^-(\omega)}{\omega-E}\ ,
\\
{\rm Im}\ \psi_{\ell\ell_3}^-(E) &=  - \frac{1}{\pi}\ {\rm P} \int_{-\infty}^\infty d\omega\ \frac{{\rm Re}\ \psi_{\ell\ell_3}^-(\omega)}{\omega-E}\ .
\end{align}
\label{eq:obsdisp}
\end{subequations}
	Note that Eqs.~\eqref{eq:obsdisp} and~\eqref{eq:statedisp} are the complex conjugate of one another;
	this fact expresses a mathematical symmetry that holds among the Hardy functions: the complex conjugate of a Hardy function from above is a Hardy function from below and vice versa,
\begin{align}
\overline{\psi^-_{\ell\ell_3}}(E) \in \mathcal{S}_-(\mathbb{R}_+) \longleftrightarrow \psi^-_{\ell\ell_3}(E) \in \mathcal{S}_+(\mathbb{R}_+)
\label{eq:cchardy}
.
\end{align}

	By complete analogy to the case of the state vector, one constructs for observable $\psi$ the RHS of Hardy functions from above denoted by $\Phi_+\subset\mathsf{H}\subset\Phi_+^\times$~\cite{ref:gadella89}, and we impose the boundary condition as
\begin{align}
\ket{\psi^-} \in \Phi_+ \subset \mathsf{H}\quad\mbox{and}\quad \ket{E\,\ell\,\ell_3\,^-} \in \Phi_+^\times \supset \mathsf{H} 
\label{eq:hardybc3}
.
\end{align}
%
%so that the wave function~\eqref{eq:obswave} is given by a bracket $\psi^-_{\ell\ell_3}(E)\equiv\braket{^-E\,\ell\,\ell_3}{\psi^-}$. 
	Here the basiskets satisfy the eigenvalue equations\footnote{The Hamiltonian in Eq.~\eqref{eq:eigenenergy-} is $H_+^\times$.}
\begin{subequations}
\begin{align}
H\,\ket{E\,\ell\,\ell_3\,^-} &= E\, \ket{E\,\ell\,\ell_3\,^-},\quad 0<E<\infty
\label{eq:eigenenergy-}
\\
\bm{L}^2 \,\ket{E\,\ell\,\ell_3\,^-} &= \ell(\ell+1)\,\ket{E\,\ell\,\ell_3\,^-},\quad \ell=0,1,2,\cdots
\\
L_3\,\ket{E\,\ell\,\ell_3\,^-} &= \ell_3\,\ket{E\,\ell\,\ell_3\,^-},\quad \ell_3=-\ell, -\ell+1, \cdots, \ell
,
\end{align}
\label{eq:hardyket-}
\end{subequations}
with a normalization
\begin{align}
\braket{^-E'\ell'\ell_3'}{E\,\ell\,\ell_3\,^-}=\delta(E'-E)\,\delta_{\ell'\ell}\,\delta_{\ell'_3 \ell_3}
\label{eq:deltanormobs}
,
\end{align}
so that $\ket{\psi^-}$ can be written as
\begin{align}
\ket{\psi^-} = \sum_{\ell\ell_3} \int_0^\infty dE\ \ket{E\,\ell\,\ell_3\,^+}\,\psi^-_{\ell\ell_3}(E)
\label{eq:obspartialwave-}
.
\end{align}
	In the Hardy RHS~\eqref{eq:hardybc3}, the one-to-one correspondence holds between the observable $\psi$ and its energy wave function $\psi^-_{\ell\ell_3}(E)\equiv\braket{^-E\,\ell\,\ell_3}{\psi^-}$ as
\begin{align}
\ket{\psi^-} \in \Phi_+ \longleftrightarrow \psi_{\ell\ell_3}^-(E) \in \mathcal{S}_+(\mathbb{R}_+)
\label{eq:hardybc4}
.
\end{align}
	From this, by following the same argument as in Eq.~\eqref{eq:calc1} to Eq.~\eqref{eq:causalstate+}, time evolution of the observable $\psi$ in the Heisenberg picture is obtained as
\begin{align}
\ket{\psi^-(t)} &= e^{iHt}\,\ket{\psi^-} \quad\mbox{for $0\leq t<\infty$ only}
\label{eq:causalstate-}
.
\end{align}
	With this vector, we have for the observable $\psi$
\begin{align}
\mathcal{O}_\psi(t) = \ketbra{\psi^-(t)}{\psi^-(t)} \quad \mbox{for $0\leq t<\infty$}
\label{eq:hardyobs}
,
\end{align}
that is exactly the semigroup time evolution~\eqref{eq:causalobs}.

	A physical interpretation of the wave function $\psi^-_{\ell\ell_3}(E)$ is such that its square-modulus, $\abs{\psi^-_{\ell\ell_3}(E)}^2$, describes energy-and-angular resolution function of the registration apparatus.
	For example, if the registration apparatus is to select a microsystem within the energy resolution function $g(E)$ of Lorentzian with a peak energy at $a'$ and the FWHM $b'$,
\begin{align}
g(E) = \frac{\sum_{\ell\ell_3}\abs{C'_{\ell\ell_3}}^2}{(E-a')^2+(b'/2)^2}
,
\end{align}
then its corresponding energy wave function is given by
\begin{align}
\psi_{\ell\ell_3}^-(E) = \frac{C'_{\ell\ell_3}}{E-(a'-ib'/2)}
\ \in\mathcal{S}_+(\mathbb{R}_+)
\label{eq:lorentzianobs}
,
\end{align}
where $C'_{\ell\ell_3}$ is determined to satisfy the normalization
\begin{align}
||\psi^-||^2=\braket{\psi^-}{\psi^-}=\int_0^\infty dE\, g(E) = 1.
\end{align}

\begin{figure}[htbp]
\begin{center}
\includegraphics*[height=2.5cm]{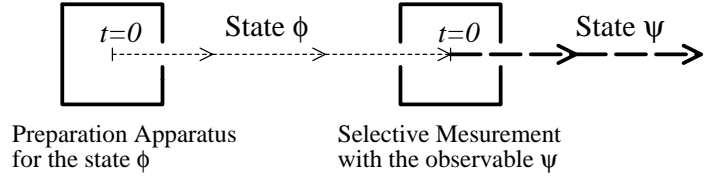}
%figuresize=46x13
\end{center}
\caption{Preparation of the state $\psi$ with the observable $\psi$.}
\label{fig:statepsi}
\end{figure}

	In the process of selective measurement, a new state can be prepared. 
	It is generally assumed in quantum mechanics that immediately after the measurement of an observable the microsystem will be in a state that has been prepared by this measurement~\cite{ref:bohm01book}.
	This means that the state prepared by the observable $\psi$ has the same energy-angular distribution as $\abs{\psi^-_{\ell\ell_3}(E)}^2$, but its state vector is in the Hardy space $\Phi_-$.
	Such a state, the {\it state} $\psi$, is uniquely obtained by the mathematical symmetry~\eqref{eq:cchardy} between Hardy functions as
\begin{align}
\ket{\psi^+} = \sum_{\ell\ell_3} \int_0^\infty dE\, \ket{E\,\ell\,\ell_3\,^+} \psi^+_{\ell\ell_3}(E)
,
\end{align}
with
\begin{align}
\psi^+_{\ell\ell_3}(E)\equiv\braket{^+E\ell\,\ell_3}{\psi^+} = \overline{\psi^-_{\ell\ell_3}}(E) \ \in\mathcal{S}_-(\mathbb{R}_+)
.
\end{align}
	In the case of Eq.~\eqref{eq:lorentzianobs}, for example, this wave function is given by
\begin{align}
\psi^+_{\ell\ell_3}(E) = \frac{\overline{C'_{\ell\ell_3}}}{E-(a'+ib'/2)}
.
\end{align}
	Thus, as shown in Fig.~\ref{fig:statepsi}, the microsystem originally prepared in the state $\phi$ ``jumps'' into the state $\psi$ due to the measurement of the observable $\psi$.
%%%
	This preparation of state brings one a new ensemble of beginnings of time.
	In an idealized situation that the time it takes for a measurement is negligible, this beginnings of time would coincide with the ensemble of ``ends of time''~\eqref{eq:endoftime} at which the eigenvalue 1 (`affirmative') of the observable $\mathcal{O}_\psi$ is registered,
\begin{align}
t=0\ : \ \left\{T'_1, T'_2, \cdots , T'_N \right\} \ \mbox{for the state $\psi$}
.
\label{eq:begoftimepsi}
\end{align}
	From this $t=0$ the time evolution of $\ket{\psi^+}$ then begins.

	To summarize, the time asymmetric boundary condition is a pair of Hardy RHSs, one for prepared states,
\begin{subequations}
\begin{align}
&\ket{\phi^+}\in\Phi_-\subset\mathsf{H}\subset\Phi_-^\times \ni \ket{E\,\ell\,\ell_3\,^+},
\intertext{and the other for observables of selective measurment,}
&\ket{\psi^-}\in\Phi_+\subset\mathsf{H}\subset\Phi_+^\times \ni \ket{E\,\ell\,\ell_3\,^-}
.
\end{align}
\label{eq:tabc}
\end{subequations}
	Although the vector spaces for states and those for observables are different, a scalar product (bracket) between their vectors, such as  $\braket{\psi^-}{\phi^+}$ or even $\braket{\psi^-}{E\,\ell\,\ell_3\,^+}$, is well defined. 
	This is because scalar product is already defined in a linear-scalar-product space, or pre-Hilbert space, from which each of the vector spaces in RHS is obtained by completion with different topology (conversion of infinite sequence)~\cite{ref:gelfand64, ref:gadella89}.
	One is thus allowed to take a scalar product between {\it any} of the elements among the Hardy RHS~\eqref{eq:tabc}.

	The scalar product of a physical importance is the transformation function between basiskets of $\Phi_+^\times$ and $\Phi_-^\times$:
\begin{align}
\braket{^-E'\ell'\ell_3'}{E\,\ell\,\ell_3\,^+} = S_{\ell\ell_3}(E)\,\delta(E'-E)\,\delta_{\ell'\ell}\,\delta_{\ell_3'\ell_3}
\label{eq:smatrix}
.
\end{align}
	Here the function $S_{\ell\ell_3}(E)$ is the S-matrix element~\cite{ref:eden66}.
	It is this function that characterizes the experiment with the state $\phi$ and the observable $\psi$.
%	This enables one to express transition probability in terms of S-matrix element.
	The transition probability~\eqref{eq:tranprob} with the time asymmetric boundary condition~\eqref{eq:tabc} is given by
\begin{align}
\mathcal{P}(t) &= \bra{\phi^+(t)}\mathcal{O}_\psi\ket{\phi^+(t)} = \abs{\braket{\psi^-}{\phi^+(t)}}^2 \quad \mbox{in the Schr\"odinger picture},\nonumber\\
			   &= \ \ \bra{\phi^+}\mathcal{O}_\psi(t)\ket{\phi^+} \ \, = \abs{\braket{\psi^-(t)}{\phi^+}}^2 \quad \mbox{in the Heisenberg picture},\nonumber\\
			   &\equiv \ \ \abs{a(t)}^2\quad \mbox{for $0\leq t<\infty$},
\end{align}
where $a(t)\equiv\braket{\psi^-}{\phi^+(t)}=\braket{\psi^-(t)}{\phi^+}$ is the time-dependent transition amplitude.
%	Now we define an S-matrix element $S_{\ell\ell_3}(E)$ as 
	With the S-matrix element defined by Eq.~\eqref{eq:smatrix}, the transition amplitude is given by
\begin{align}
a(t)=\sum_{\ell\ell_3}\int_0^\infty dE\ e^{-iEt}\,\overline{\psi^-_{\ell\ell_3}}(E)\phi^+_{\ell\ell_3}(E)\, S_{\ell\ell_3}(E)
\label{eq:amp}
,
\end{align}
where Eqs.~\eqref{eq:hardyket+}--\eqref{eq:statepartialwave+}~and Eqs.~\eqref{eq:hardyket-}--\eqref{eq:obspartialwave-} have been used.
	Due to the exponential factor $e^{-iEt}$ bounded for $0\leq t<\infty$, this integral can converge in the lower-half complex (energy) plane, where the wave functions $\overline{\psi^-_{\ell\ell_3}}(E)$ and $\phi_{\ell\ell_3}^+(E)$ have no singularities (by Eq.~\eqref{eq:cchardy} they are both the Hardy function from below).
	The transition amplitude $a(t)$ is therefore determined by singularities, e.g., poles, of the S-matrix element $S_{\ell\ell_3}(E)$ in the lower-half plane.
	Thus the S-matrix elements are responsible for the dynamics of microsystem~\cite{ref:bohm97}.

%%%%%%%%%%%%%%%%%%%%%%%%%%%%%%%%%%%%%%%%%%%%%%%%%%%%%%%%%%%%%
\section{Conclusion~\label{sec:conclusion}}
%%%%%%%%%%%%%%%%%%%%%%%%%%%%%%%%%%%%%%%%%%%%%%%%%%%%%%%%%%%%%\
	From the preparation-registration arrow of time, we have obtained the lower-bounded time parameter $t$ which represents an ensemble of time intervals.
	This has lead to semigroup time evolutions of states and observables.
	The semigroups have been achieved by choosing time-asymmetric boundary conditions, a pair of Hardy rigged Hilbert spaces (RHSs) for both Sch\"odinger and Heisenberg pictures, to integrate the dynamical equations.
	The energy wave functions in those Hardy RHSs satisfy the dispersion relations and their complex conjugate.

	The S-matrix elements have been defined as transformation functions between two functional spaces of Hardy RHSs.
	The lower-boundedness of $t$ suggests that the S-matrix elements have singularities, if any, in the lower-half of complex energy plane.

\section{Acknowledgment}
Author (YS) would like to thank M.\ Gadella for his valuable suggestions.
The research was partially supported by USNSF Award No OISE-0421936.

\appendix
%%%%%%%%%%%%%%%%%%%%%%%%%%%%%%%%%%%%%%%%%%%%%%%%%%%%%%%%%%%%%
\section{Hardy functions}
%%%%%%%%%%%%%%%%%%%%%%%%%%%%%%%%%%%%%%%%%%%%%%%%%%%%%%%%%%%%%
	In this appendix, we briefly present some of the properties of Hardy functions~\cite{ref:hardyfunction} relevant to this article.

	The Hardy function from above $h_+$ and the from below $h_-$ are analytic functions satisfying the following square-integrability:
\begin{align}
\int_{-\infty}^{\infty} d\omega\, \abs{h_\pm(\omega \pm i\gamma)}^2=k <\infty 
\quad\mbox{for any fixed $\gamma >0$}
\label{eq:defhardy}
,
\end{align}
where $k$ depends on a particular form of the function $h_\pm$.
	This is necessary and sufficient condition for the Hilbert transform to hold\footnote{See Sec.\ 10.2 of Ref.~\cite{ref:goldberger64}.},
\begin{subequations}
\begin{align}
{\rm Re}\ h_\pm(\omega) &= \pm \frac{1}{\pi}\ {\rm P} \int_{-\infty}^\infty d\omega'\ \frac{{\rm Im}\ h_\pm(\omega')}{\omega'-\omega}
\ ,
\\
{\rm Im}\ h_\pm(\omega) &= \mp \frac{1}{\pi}\ {\rm P} \int_{-\infty}^\infty d\omega'\ \frac{{\rm Re}\ h_\pm(\omega')}{\omega'-\omega}
\ .
\end{align}
\label{eq:dispersion}
\end{subequations}
	This relation explicitly shows that a Hardy function has both nonzero real and imaginary parts, i.e., cannot take a pure real or pure imaginary value, otherwise zero on the entire real line.
	The complex conjugate of the Hardy function from above (below) is the Hardy function from below (above), i.e., $\overline{h_\pm} \in \mathcal{S}_\mp(\mathbb{R}_+)$.

	One can generate hardy functions out of a square-integrable functions $f(t)$ which vanish for negative values of $t$, i.e., $f(t)=\theta(t)f(t)$ holds, 
\begin{align}
||f||^2=\int_{-\infty}^\infty dt\ \abs{f(t)}^2
=
\int_{0}^\infty dt\ \abs{f(t)}^2 < \infty
.
\end{align}
	The Paley-Wiener theorem of Hardy function states that Hardy function from above $h_+(\omega)$ on the real axis is obtained from the Fourier transform of $f(t)$ as
\begin{align}
h_+(\omega) 
&= \int_{-\infty}^\infty dt\ e^{i \omega t} f(t)
= \int_0^\infty dt\ e^{i \omega t} f(t)
\label{eq:pw}
.
\end{align}
	The Hardy function from above is quite common in signal processing and physics, and sometimes referred to as {\it causal transform}~\cite{ref:toll56, ref:nussenzveig72}.
	From Eq.~\eqref{eq:pw}, Hardy function from below is immediately obtained by taking its complex conjugate as $h_-(\omega)=\overline{h_+(\omega)}$. 
	By the Titchmarsh theorem of Hardy function, these functions are guaranteed to be analytic, in the upper half plane for $h_+$ and in the lower half plane for $h_-$, as
\begin{align}
h_\pm(\omega \pm i\gamma) = \pm\frac{1}{2\pi i}\int_{-\infty}^\infty d\omega'\ \frac{h_\pm(\omega')}{\omega'-(\omega \pm i\gamma)}
\ \mbox{for $\gamma>0$}.
\end{align}
%
%	All the above integrals are done by Lebesgue measure, but, in physics, we mostly depend on Riemann integrals and that is enough to describe observed phenomena.
%	So we take not the spaces of the mathematical Hardy functions $\mathcal{H}_\pm$ but their subset where Riemann integral is always available, $\Phi_\pm$, as $\Phi_\pm \subset \mathcal{H}_\pm$, called Hardy Schwartz spaces.
%	On this account our hardy functions are denoted by~\cite{ref:gadella83}
%
%\begin{align}
%h_\pm = \mathcal{H}_\pm \cap \mathcal{S}|_{\mathbb{R}_+}
%.
%\end{align}
%
%	These functions are also called Hardy functions, and they are the ones we use in practice.
%		In order to make use of this boundary condition in the concrete example of free stable particle in the following section, we here give some example of the Hardy functions.	
%	With an arbitrary square-integrable function $f(t)$ which vanishes on the negative real axis, $f(t)=\theta(t)f(t)$, so that
%
%\begin{align}
%\int_{-\infty}^\infty dt\ \abs{f(t)}^2 &= \int_0^\infty dt\ \abs{f(t)}^2 < \infty
%,
%\end{align}
%
%Hardy functions $h_+(E)$ and $h_-(E)$ are obtained by its Fourier and inverse Fourier transforms, respectively, as
%
%\begin{align}
%h_\pm(E) &= \int_{-\infty}^\infty dt\ e^{\pm iE\,t}\,f(t)
%.
%\end{align}
%
	Thus one obtains various Hardy functions by performing the integral~\eqref{eq:pw} for various $f(t)$.
	Following are two simple examples:
\begin{itemize}
\item
For $f(t) = \theta(t)\, e^{i(a+ib)\,t}$, one obtains
\begin{align}
h_\pm(\omega) = \frac{\pm i}{(a+ib)\pm \omega}\quad\mbox{for $b>0$ and $a$ real}.
\end{align}
The square modulus of this Hardy function is a Lorentzian function.
% 
%
%\item
%Hardy functions form the real part of above $f(t)$:
%
%\begin{align}
%f(t) &= \theta(t)\, e^{-bt}\cos{(at)}
%\longrightarrow 
%h_\pm(E) = \frac{b\mp iE}{a^2+(b\mp iE)^2}\quad\mbox{for $b>0$}
%\end{align}
\item
For $f(t) = \theta(t)\, e^{-b\,t}\sin{(a\,t)}$, one obtains
\begin{align}
h_\pm(\omega) = \frac{a}{a^2+(b\mp i\omega)^2}\quad\mbox{for $b>0$ and $a$ real}.
\end{align}
	This Hardy function has been used in classical electrodynamics to describe the propagation of light in a dispersive medium~\cite{ref:jackson98}, where the physics terminology ``dispersion relations'' for Eq.~\eqref{eq:dispersion} is originated in.
\end{itemize}
%

%%%%%%%%%%%%%%%%%%%%%%%%%%%%%%%%%%%%%%%%%%%%%%%%%%%%%%%%%%%%%
%\section{}
%%%%%%%%%%%%%%%%%%%%%%%%%%%%%%%%%%%%%%%%%%%%%%%%%%%%%%%%%%%%%

%%%%%%%%%%%%%%%%%%%%%%%%%%%%%%%%%%%%%%%%%%%%%%%%%%%%%%%%%%%%%%

%%%%%%%%%%%%%%%%%%%%%%%%%%%%%%%%%%%%%%%%%%%%%%%%%%%%%%%%%%%%%%%%

%% The Appendices part is started with the command \appendix;
%% appendix sections are then done as normal sections
%% \appendix

%% \section{}
%% \label{}

%% References
%%
%% Following citation commands can be used in the body text:
%% Usage of \cite is as follows:
%%   \cite{key}          ==>>  [#]
%%   \cite[chap. 2]{key} ==>>  [#, chap. 2]
%%   \citet{key}         ==>>  Author [#]

%% References with bibTeX database:

%\bibliographystyle{model1a-num-names}
%\bibliography{<your-bib-database>}

\begin{thebibliography}{99}

\bibitem{ref:kronig26}
%[kronig26]
R.\ de Kronig, J.\ Opt.\ Soc.\ Am.\ {\bf 12}, 547 (1926).


\bibitem{ref:kramers27}
%[kramers27] 
H.\ A.\ Kramers, Atti.\ congr.\ intern.\ fisici Como {\bf 2}, 545 (1927). 
%{\it Estratto dagli Atti del Congresso Internazionale de Fisici Como} (Nicolo Zonichelli, Bologna, 1927); {\it Collected Scientific papers} (North-Holland, Amsterdam, 1956) 


\bibitem{ref:toll56}
%[toll56] 
J.\ S.\ Toll, Phys.\ Rev. {\bf 104}, 1760 (1956).




\bibitem{ref:kronig42}
%[kronig42] 
R.\ Kronig, Ned.\ Tijdschr.\ Natuurk {\bf 9}, 402 (1942).





\bibitem{ref:nussenzveig72}
%[nussenzveig72] 
H.\ M.\ Nussenzveig, {\it Causality and Dispersion Relations} (Academic Press, New York, 1972).



\bibitem{ref:schutzer51}
%[schutzer51] 
W.\ Schutzer and J.\ Tiomno, Phys.\ Rev.\ {\bf 83}, 249 (1951).



\bibitem{ref:vankampen53II}
%[vankampen53II] 
N.\ G.\ van Kampen, Phys.\ Rev.\ {\bf 91}, 1267 (1953).



\bibitem{ref:wigner55}
%[wigner55] 
E.\ P.\ Wigner, Amer.\ J.\ Phys.\ {\bf 23}, 371 (1955).


\bibitem{ref:gell-mann54}
%[gell-mann54] 
M.\ Gell-mann, M.\ L.\ Goldberger, and W.\ Thirring, Phys.\ Rev.\ {\bf 95}, 1612 (1954).


\bibitem{ref:weinberg95}
%[weinberg95] 
S.\ Weinberg, 
{\it The Quantum Theory of Fields} Vol.1 (Cambridge University Press, Cambridge, 1995), and references thereof.




\bibitem{ref:ludwig85}
%[ludwig85]
G.\ Ludwig, {\it Foundations of Quantum Mechanics}, Vol.\ I and II (Springer-Verlag, Berlin, 1985).



\bibitem{ref:harshman98}
%[harshman98]
A.\ Bohm and N.\ L.\ Harshman in 
A.\ Bohm, H.\ Doebner, and P.\ Kielanowski Ed., {\it Irreversibility and causality}, Lecture Notes in Physics Vol.\ 504, page 181 (Springer, Berlin Heidelberg, 1998).


\bibitem{ref:dirac58}
%[dirac58]
P.\ A.\ M.\ Dirac, 
{\it The Principles of Quantum Mechanics} (Oxford University Press, Oxford, 1958), 4th Ed.



\bibitem{ref:bohm78}
%[bohm78] 
A.\ Bohm, Lett.\ Math.\ Phys.\ {\bf 3} 455 (1978); 
J.\ Math.\ Phys.\ {\bf 53}, 2813 (1981). 


\bibitem{ref:bohm97}
%[bohm97]
A.\ Bohm, S.\ Maxson, M.\ Loewe, and M.\ Gadella, Physica A {\bf 236}, 485 (1997)





\bibitem{ref:neumann55}
%[neumann55]
J. von Neumann, {\it Mathematical Foundations of Quantum Mechanics} (Princeton University Press, Princeton, 1955).

\bibitem{ref:wightman00}
%[wightman00]
R.\ F.\ Streater and A.\ S.\ Wightman, {\it PCT, Spin and Statistics, and All That} (Princeton University Press, Princeton, 2000)



\bibitem{ref:stone-vonneumann}
%[stone-vonneumann]
M.\ H.\ Stone, Ann.\ Math.\ {\bf 33}, 643 (1932);
J.\ von Neumann, Ann.\ Math.\ {\bf 33}, 567 (1932).

\bibitem{ref:bohm06}
%[bohm06] 
A.\ Bohm, P.\ Kielanowski, and S.\ Wickramasekara, 
Ann.\ Phys.\ {\bf 321}, 2299 (2006).


\bibitem{ref:ionexp}
%[ionexp]
H.\ Dehmelt, Bull.\ Am.\ Phys.\ {\bf 20}, 60 (1975);
W.\ Nagourney, J.\ Sandberg, H.\ Dehmelt, Phys.\ Rev.\ Lett.\ {\bf 56}, 2797 (1986). 


%\bibitem{ref:scattering}
%[scattering]
\bibitem{ref:goldberger64}
%[goldberger64]
M.\ L.\ Goldberger and K.\ M.\ Watson, 
{\it Collision Theory} (Wiley, New York,1964)
.




\bibitem{ref:newton82}
%[newton82] 
R.\ G.\ Newton, 
{\it Scattering Theory of Waves and Particles} 2nd ed.\ (Springer-Verlag, New York, 1982); (Dover, New York, 2002)
.



\bibitem{ref:hegerfeldt94}
%[hegerfeldt94]
G.\ C.\ Hegerfeldt, Phys.\ Rev.\ Lett.\ {\bf 72}, 596 (1994);
G.\ C.\ Hegerfeldt in A.\ Bohm, H.\ Doebner, and P.\ Kielanowski Ed., Lecture Notes in Physics Vol.\ 504, page 238 {\it Irreversibility and causality} (Springer, Berlin Heidelberg, 1998).

\bibitem{ref:gelfand64}
%[gelfand64]
J.\ M.\ Gel'fand and N.\ Ya Vienkin, 
{\it Generalized Functions} Vol.4 (Academic Press, New York, 1964); 
K.\ Maurin, {\it Generalized Eigenfunction Expansions and Unitary Representations of Topological Groups} (Polish Scientific Publishers, Warsaw, 1968).

\bibitem{ref:hardyfunction}
%[hardyfunction]
P.\ L.\ Duren, {\it $H^P$ Spaces} (Academic Press, New York, 1970);
K.\ Hoffman, {\it Banach Spaces of Analytic Functions} (Prentice-Hall, Englewood Cliffs, NJ, 1962; Dover, New York, 1988).



\bibitem{ref:vanwinter}
%[vanwinter]
%[vanwinter71]
C.\ van Winter, Trans.\ Am.\ Math.\ Soc.\ {\bf 162}, 103 (1971)
;
%[vanwinter74]
J.\ Math.\ Anal.\ Appl.\ {\bf 47}, 633 (1974)
.



\bibitem{ref:jackson98}
%[jackson98] 
J.\ D.\ Jackson, 
{\it Classical Electrodynamics} 3rd ed. (John Wiley \& Sons, Singapore, 1998)


\bibitem{ref:gadella89}
%[gadella89]
A.\ Bohm and M.\ Gadella, 
{\it Dirac Kets, Gamow Vectors, and Gel'fand Triplets} Lecture Notes in Physics Vol.348 (Springer, Berlin, 1989).


\bibitem{ref:haag96}
%[haag96]
R.\ Haag, 
{\it Local Quantum Physics: Fields, Particles, Algebras} 2nd ed.\ (Springer, Berlin, 1996).



\bibitem{ref:sakurai94}
%[sakurai94] 
J.\ J.\ Sakurai, S.\ F.\ Tuan ed.,  
{\it Modern Quantum Mechanics}  Revised ed.\ (Addison-Wesley, New York, 1994).


\bibitem{ref:schwinger59}
%[schwinger59] 
J.\ Schwinger, Proc.\ Nat.\ Acad.\ Sc.\ {\bf 45}, 1542 (1959); {\bf 46}, 257 (1960); {\bf 46}, 570 (1960).




\bibitem{ref:bohm01book}
%[bohm01book]  
A.\ Bohm, {\it Quantum Mechanics: Foundations and Applications} (Springer, New York, 2001), 3rd printing. 




\bibitem{ref:eden66}
%[eden66]
R.\ J.\ Eden, P.\ V.\ Landshoff, D.\ I.\ Olive, and J.\ C.\ Polkinghorne, 
{\it The Analytic S-matrix} (Cambridge University Press, Cambridge, 1966).


%%%%%%%%%%%%%%%%%%%%%%%%%%%%%%%%%%%%%%%%%%%%%%%%%%%%%%%%%%%%%%%%
\end{thebibliography}

%% Authors are advised to submit their bibtex database files. They are
%% requested to list a bibtex style file in the manuscript if they do
%% not want to use model1a-num-names.bst.

%% References without bibTeX database:

% \begin{thebibliography}{00}

%% \bibitem must have the following form:
%%   \bibitem{key}...
%%

% \bibitem{}

% \end{thebibliography}

\end{document}